\documentclass[12pt,a4paper]{amsart}

\usepackage[american]{babel}
\usepackage{bm,amsmath,amssymb}
\usepackage{amsbsy,amsfonts,array}
\usepackage{verbatim,graphicx}
\usepackage{epstopdf}
\usepackage{subfigure}
\graphicspath{{./immagini/}}
\usepackage{color}
\DeclareGraphicsExtensions{.jpg,.png,.pdf}
\usepackage{dsfont} 
\usepackage{amsthm}
\usepackage{textcomp}                      
\usepackage[mathscr]{euscript}

\usepackage[margin=3cm]{geometry}
\usepackage[P,R]{dblfont}

\newcommand{\be}{\begin{equation*}}
\newcommand{\ee}{\end{equation*}}
\newcommand{\bcr}{\begin{center}}
\newcommand{\ecr}{\end{center}}
\newcommand{\beq}{\begin{equation}}
\newcommand{\eeq}{\end{equation}}
\newcommand{\Frac}{\displaystyle\frac}

\newcommand{\fl}{\textrm{fl}}
\newcommand{\s}{\textrm{s}}

\newcommand{\e}{\textrm{e}}
\newcommand{\n}{\textrm{n}}
\newcommand{\q}{\textrm{q}}
\renewcommand{\v}{\textrm{v}}
\newcommand{\ECM}{\textrm{ECM}}
\newcommand{\cells}{\textrm{cells}}

\newcommand{\qui}{\textrm{qui}}
\newcommand{\apo}{\textrm{apo}}
\newcommand{\cell}{\textrm{cell}}

\definecolor{darkmagenta}{rgb}{0.55, 0.0, 0.55}
\newcommand{\rs}[1]{\textcolor{black}{#1}}

\newtheorem{assumption}{Assumption}[section]

\begin{document}

\title[A Poroelastic Mixture Model, Part I]
{A Poroelastic Mixture Model of Mechanobiological
Processes in Tissue Engineering. \\
Part I: Mathematical Formulation}

\author{Chiara Lelli$^{1}$ \and Riccardo Sacco$^{1}$ 
\and Paola Causin$^{2}$
\and Manuela T. Raimondi$^{3}$}

\address{$^{1}$ Dipartimento di Matematica, Politecnico di Milano, \\
Piazza Leonardo da Vinci 32, 20133 Milano, Italy \\
$^{2}$ Dipartimento di Matematica \lq\lq F. Enriques\rq\rq,
Universit\`a degli Studi di Milano, \\
via Saldini 50, 20133 Milano, Italy \\
$^{3}$ Dipartimento di Chimica, Materiali e Ingegneria
Chimica \\ \lq\lq Giulio Natta\rq\rq, Politecnico di Milano, 
Piazza Leonardo da Vinci 32, 20133 Milano, Italy}

\date{\today}

\begin{abstract}
\rs{An adequate control of cell response in tissue engineering 
applications is of utmost importance to obtain products suitable to clinical practice. This paper is the first part of a series of 
two connected publications in which we study via mathematical 
tools the cultivation in bioreactors of articular chondrocytes. 
The proposed model combines poroelastic theory of mixtures and 
cellular population models into a framework including stress state and 
oxygen tension as main determinants of engineered culture evolution.
The special mechanosensitivity of articular chondrocytes to the surrounding 
environment is accounted for in the model through the novel concept of "force isotropy" acting on the cell which is assumed as 
the promoting factor of the production of new cells or extracellular matrix.}
\end{abstract}

\maketitle


\vspace*{5pt}

{\bf Abbreviations:} TE (tissue engineering); ECM (extracellular matrix);
ACC (\rs{articular chondrocyte cell}).

\section{Introduction}
\label{intro}
The poor intrinsic healing capacity of articular cartilage highlights the strong clinical need for reparative therapies.
While existing techniques with autograft/allograft transplants have had limited success, \rs{artificially engineered} 
cartilage offers a possible alternative. Cartilage \rs{TE}
typically utilizes the seeding of ACCs
into polymeric scaffolds. In this arrangement, the ECM
concentration is designed to gradually increase and form neo-tissue.
Despite the fact that sophisticated
scaffold architectures have been devised to optimize ACC growth~\cite{Langer1993,Novakovic,Freed2000,Martin2004,Raimondi2005},
current culture methodologies present a bottle-neck by
being inefficient and sub-optimal. In particular, the cultivation of
ACCs is well known to be
bioprocess-dependent. This fact is exemplified by the
extensive apoptosis of internal regions
observed in 3D cellularised constructs
maintained in static culture, inside which nutrient levels are well below a critical level~\cite{Novakovic,Freed2000}.
Moreover,  limitations exist in maintaining ACCs in cartilaginous ECM-forming state after several passages of monolayer expansion culture~\cite{raimondipellets2011,Raimondi2006a}.
As a matter of fact, ACCs have the potential to proliferate
in culture and to express their mature phenotype and, consequently,
synthesize cartilaginous ECM~\cite{Freed,Sengers}, but these conditions
coexist in the cell culture in a
delicate dynamical equilibrium~\cite{Nikolaev}. Such an equilibrium is affected by
extrinsic cues, including soluble growth factor signaling, culture substrates, oxygen tension, nutrient/metabolite content and
pH (see~\cite{Nava2012} and the references cited therein).
Furthermore, experimental data indicate that early-passage human chondrocytes adapted to fluid-dynamically stimulated culture conditions exhibit
differences in transcriptional profiles, growth and matrix
deposition compared to statically cultured cells~\cite{Nikolaev,Osborne}.
More in general, mechanical
signals are recognized to have a deep influence on ACC behavior both in vivo and in vitro
(see~\cite{Bader2011} and references therein).

\begin{figure}[h!]
\centering
\includegraphics[width=0.7\textwidth]{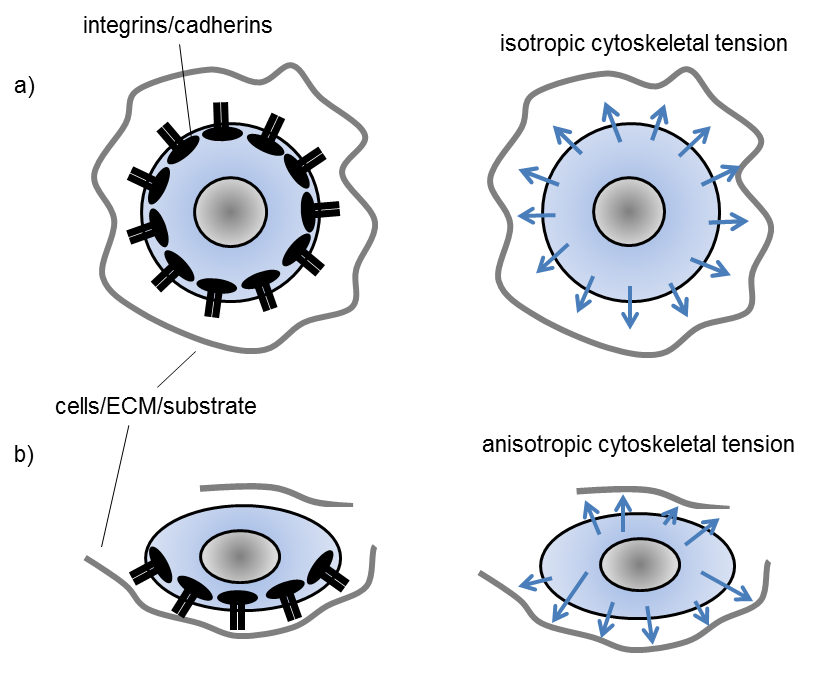}
\caption{Concept of isotropic/anisotropic stress state of a cell.}
\label{fig:iso_aniso_growth}
\end{figure}

In this work, which is the first part of a series of two connected publications,
we focus on the conceptual mechanobiological framework developed
in~\cite{Nava2012,Nava2014}, where the idea of ``force isotropy'' on the cell
is introduced. Namely, the complex system of traction forces exerted by the cell
on the surrounding environment -substrate, ECM, other cells- induces cytoskeletal
tensional states capable of triggering signalling transduction
cascades regulating functional cell
behavior, for example, the \rs{import flow} of specific
transcription factors in the nucleus.
If cell adhesion-mediated traction forces have
similar magnitude at varying orientations over the cell surface, the cell nucleus tends to maintain a roundish
morphology and this condition is defined as ``isotropic cytoskeletal
tension'' (see Fig.~\ref{fig:iso_aniso_growth}(a)). On the contrary, if cell adhesion-mediated traction forces
have different
magnitudes at varying spatial orientations, then the cell nucleus tends to elongate and this condition is defined as ``anisotropic cytoskeletal
tension'' (see Fig.~\ref{fig:iso_aniso_growth}(b)).

\rs{To investigate these conditions,} 
due to the continuum-based approach adopted throughout this paper, we do not describe the complex system of adhesion forces exerted on a single cell;
rather, we use the isotropic/anisotropic parts of the local stress tensor associated with the biological construct to mathematically represent the isotropic/anisotropic cell adhesion state. To this purpose, we introduce the following qualitative definitions:
if the an\-i\-so\-tro\-pic part of the local stress 
tensor is lower than a fixed threshold then the local stress state 
of the system is
\emph{isotropic} otherwise the local stress state of the
system is \emph{anisotropic}. We refer to Sect.~\ref{sec:production_terms}
for a quantitative characterization of these concepts.

\begin{figure}[bt!]
\centering
\includegraphics[width=0.7\textwidth]{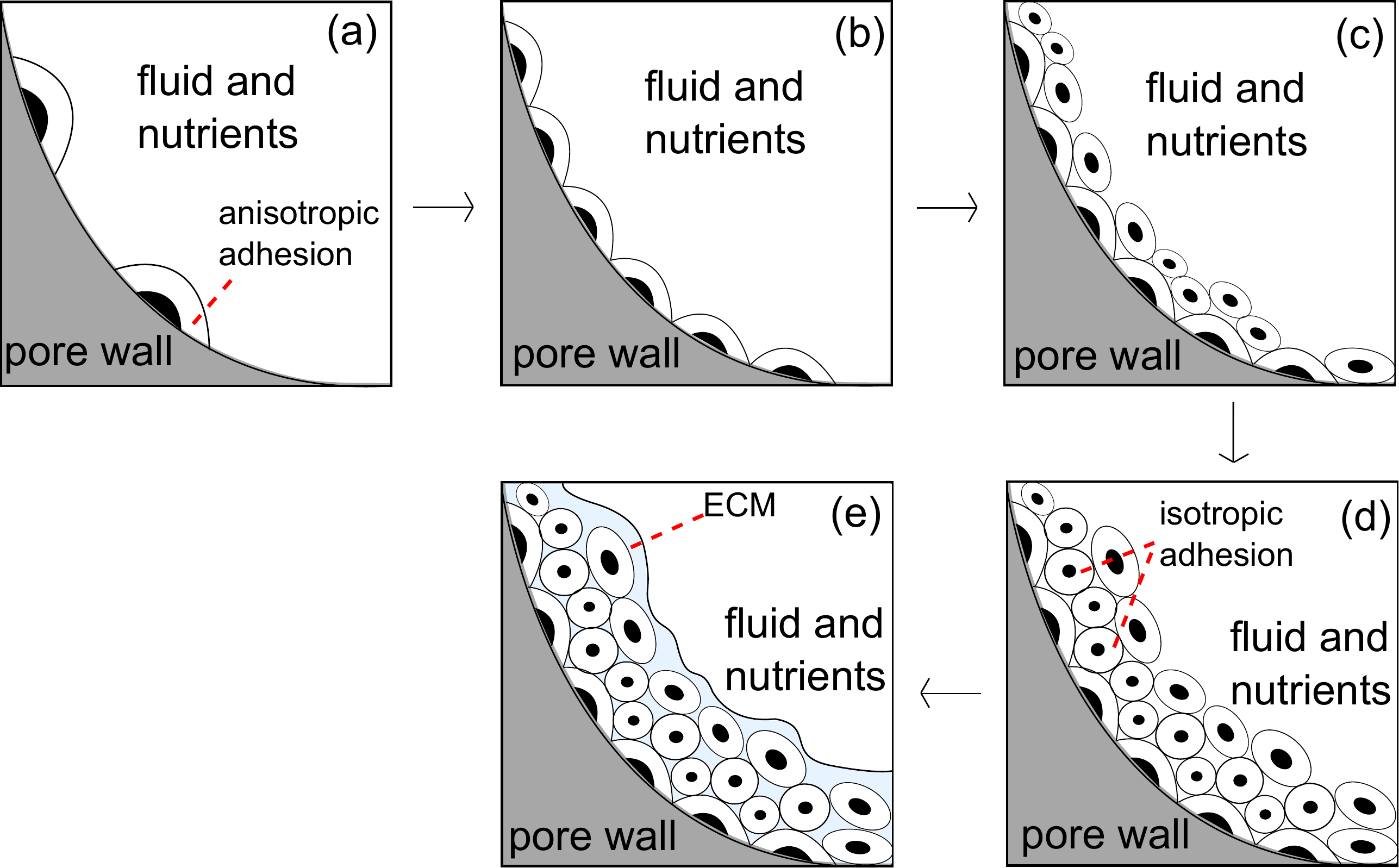}
\caption{Various phases of tissue growth inside a scaffold pore:
a) seeding phase and cell polarization; b) proliferation and
formation of a monolayer;
c), d) formation of new construct layers; e) ECM secretion.}
\label{fig:cellAdhesion}
\end{figure}

How a colony of ACCs seeded in a 3D porous scaffold
can experience the above mechanical conditions
is schematically illustrated
in Fig.~\ref{fig:cellAdhesion}.

When cells are first seeded in the scaffold, they form
a thin layer covering the surface. Since the characteristic dimension
of the local scaffold curvature is much larger than cell size,
cells find themselves in a local planar condition
(see Fig.~\ref{fig:cellAdhesion} (a)). Exerting adhesion
forces on the scaffold surface, cells tend to assume a spread elongated
shape, orienting themselves along a preferred polarization axis.
According to the concept of ``force isotropy'', this represents
an anisotropic stress state. 
In this condition, the probability that the single cell enters into a proliferative state is enhanced \rs{(cf.~\cite{Nava2012})}.
The cell entering into the mitotic phase divides along
a polarization axis represented by the direction of its long axis.
This situation persists until all the pore surface is covered with cells (Fig.~\ref{fig:cellAdhesion} (b)). From this moment on,
cells start to occupy the empty space of the pore (Fig.~\ref{fig:cellAdhesion} (c,d)). Cells in contact with other cells sense an isotropic stress state condition,
which drives the cell towards a mature differentiated
phenotype, characterized by ECM secretion (Fig.\ref{fig:cellAdhesion} (e)).
\rs{At present, there exists limited knowledge of how different levels
of mechanical isotropy are transduced into functional cell fate.
Incorporation of this newly-hypothesized mechanobiological mechanism into
a computational tool for the prediction of tissue growth could represent a significant step forward towards a better understanding of the underlying mechanisms and consequently
to an improvement in bioprocessing methodologies.}

An outline of the presentation of theory and
results is as follows.
In Sect.~\ref{sec:multiphase_mixture} we state notation
and basic assumptions; in Sect.~\ref{sec:kinematics} we
introduce the basic definitions of the kinematical model;
in Sect.~\ref{sec:balance_laws} we
illustrate the conservation laws of the mechanobiological model.
In Sect.~\ref{sec:production_terms} we describe the mathematical models
of production terms; in Sect.~\ref{sec:bio-mechanical_models} we
conclude the formulation of our model by providing a
mathematical description of the mechanobiological processes involving cell populations and ECM secretion.
In the final Sect.~\ref{sec:conclusions_part_1} we summarize the main
modeling contributions proposed in this first part of our study on
ACCs and emphasize possible research directions that should be worth investigating in a future work.

\section{Multiphase modeling}\label{sec:multiphase_mixture}
\rs{In this section we develop 
a mathematical model} based on the representation of
the ensemble of the growing cartilaginous biomass by the
mixture theory. In this framework, equations are postulated for the balance of mass
and momentum for each constituent and then for the entire mixture
according to the following ideas:
\begin{itemize}
\item[a)] the growing biomass is treated as a mixture
composed by a multiphase solid mass and an interstitial fluid, this latter
representing a fraction of the order of
$65-80$\% in mass of the total biomass. The multiphase solid consists of ACCs and of ECM.
The cell component includes different ACC populations
(proliferative, ECM secreting, quiescent), as in the works of Sengers~\cite{Sengers2} and Ducrot~\cite{Ducrot};
\item[b)] the poroelasticity theory is used to model the interaction
of deformation and fluid flow in the fluid-saturated porous, elastic solid~\cite{biot1,Coussy};
\item[c)] the kinematics of the solid phase of the mixture is based on an in\-fi\-ni\-tesimal--deformation
approach, including the effect on the stress field of biological growth,
according to the formulation proposed by Klisch and co--authors~\cite{Klisch,Klisch0,Klisch1};
\item[d)]  the mass conservation balance for each single constituent and for the mixture are written according to the formulation introduced by Lemon and co-authors~\cite{Lemon2006,Lemon2007} and extensively analyzed
in~\cite{Tosin2,Tosin};
\item[e)] the mass exchange terms, including the rate of switch of cells from a population to the other, are tuned according to
the nutrient level, this latter being itself an unknown of the problem,
and to the stress state locally experienced by
the mixture, which may drive cells into a certain functional behavior pool.
\end{itemize}

In the following, we use the term
\lq\lq phase\rq\rq{} when we refer to the solid or to the fluid part of the
mixture, while the term \lq\lq component\rq\rq{}
is used to refer to any of the constituents of
the solid phase (cell populations and ECM). When it is not necessary to
distinguish between phase and component, we simply use
the term \lq\lq species\rq\rq.
The meaning of the subscripts used
throughout the article is as follows:
s=solid phase, fl=fluid phase, cells=cell component of the solid phase,
ECM= extracellular component of the solid phase.

We let $\mathbf{x}$ and $t$ denote the space and time variables, respectively.
We use the convention that the dependence of all
variables and model parameters on $\mathbf{x}$ and $t$
is left understood except otherwise stated.

The geometrical configuration of the mixture
is identified by the open
bounded set $\Omega \subset \mathbb{R}^d$ ($d=3$ unless otherwise
specified). The domain $\Omega$ does not evolve in time, rather,
it is the amount of each species at a point $\mathbf{x} \in \Omega$
that changes with $t$ due to cell proliferation and matrix deposition.
This is the precise sense of the concept of
``growing mixture". From now on, we denote by
$\mathcal{Q}_{T_{end}}: = \Omega \times (0, T_{end})$
the space-time cylinder
where the TE problem is studied, $T_{end} >0$ being the
final time of culture process.
\begin{figure}[h!]
\centering
\includegraphics[width=0.6\textwidth]{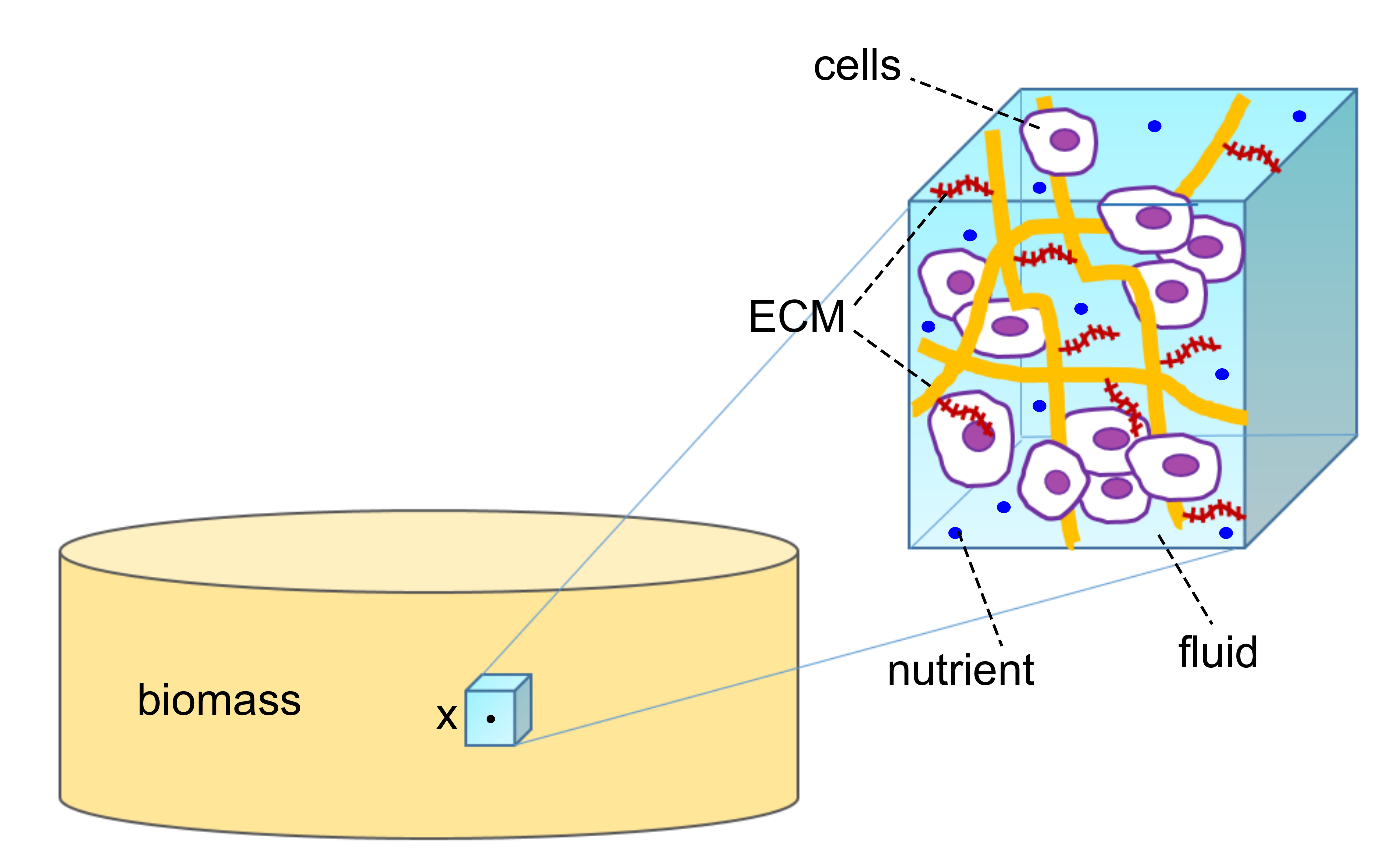}
\caption{A schematic view of the computational
domain $\Omega$ with a detailed view of a typical REV
where the various phases and components of the growing mixture are
identified.}
\label{fig:REV}
\end{figure}

Let us now introduce the basic mathematical elements for developing
a theory of the growing mixture. Referring to
Fig.~\ref{fig:REV}, for all $t>0$ we associate with a generic point
$\mathbf{x} \in \Omega$ a fixed representative elementary volume (REV)
$\mathcal{V}^\mathbf{x}$ (see~\cite{whitaker1999}) and denote by
$|\mathcal{V}^\mathbf{x}|$ its $d$-dimensional volume. Then,
for each component $i=\fl, \s$
of the growing mixture, we define the volume fraction
$$
\phi_i (\mathbf{x},t) =
\Frac{|\mathcal{V}^{\mathbf{x}}_{i}(\mathbf{x},t)|}
{|\mathcal{V}^\mathbf{x}|}
\qquad  \forall \mathbf{x} \in \Omega, \qquad \forall t >0
$$
as the time evolving ratio of the volume occupied by the $i$-th component
in the REV and the volume of the REV itself.
We further have
\begin{subequations}\label{eq:basic_foundations}
\begin{align}
& \phi_{\s}(\mathbf{x},t) =
\phi_{\cells}(\mathbf{x},t) + \phi_{\ECM}(\mathbf{x},t)
& \qquad  \forall \mathbf{x} \in \Omega, \qquad \forall t >0.
\label{eq:phi_solid}
\end{align}

According to the biochemical \rs{hypothesis a)}
we consider three ACC populations:
proliferating, ECM secreting state or quiescent state, denoted by the letters $\n$, $\v$ and $\q$, respectively.
Cells in the quiescent state may reversibly lead
to the mitotic or ECM-secreting populations or to apoptosis depending on
nutrient level. Then, we have
\begin{align}
& \phi_{\cells}(\mathbf{x},t) =
\phi_{\n}(\mathbf{x},t) + \phi_{\v}(\mathbf{x},t)  + \phi_{\q}(\mathbf{x},t)
& \qquad  \forall \mathbf{x} \in \Omega, \qquad \forall t >0.
\label{eq:phi_cells}
\end{align}
In the remainder of the article we denote by
$$
\boldsymbol{\phi} = \left[\phi_{\fl}, \,
\phi_\n, \, \phi_\v, \, \phi_\q, \, \phi_{\ECM} \right]^T
$$
the vector-valued function comprising the volume fractions of the
fluid phase, the three cellular populations and the ECM.

\medskip

\noindent The following standard assumptions on the mixture
are also considered.

\begin{assumption}[Fully saturated mixture]\label{ass:fullysaturation}
The mixture is fully saturated, i.e.
\begin{align}
& \phi_{\mathrm{s}}(\mathbf{x},t) +
\phi_{\mathrm{fl}}(\mathbf{x},t) = 1 & \qquad
\forall \mathbf{x} \in \Omega, \qquad \forall t >0.
\label{eq:saturation}
\end{align}
\end{assumption}

Relation~\eqref{eq:saturation} is referred to as
\emph{saturation condition}~\cite{Tosin,Sengers,Araujo}
and excludes the possibility of the formation of voids or
air bubbles inside the growing mixture.

\begin{assumption}[Intrinsic incompressibility]\label{ass:intrincomp}
All species  constituting the growing mixture
have the same (constant) mass density $\rho_w$
of the physiological interstitial fluid assimilated to water~\cite{Lemon2006,Tosin,Araujo, Klisch_intrinsic_incompressibility,frijns}.
\end{assumption}
Assumption~\ref{ass:intrincomp} is not, in general,
equivalent to assuming that the \emph{whole}
mixture is incompressible (see~\cite{Tosin} p.~629).

\begin{assumption}[Closed mixture]\label{ass:closedsystem}
The mixture is closed, this meaning that the system does not
exchange mass with external mass sources or sinks~\cite{Tosin}.
\end{assumption}
\end{subequations}

\section{Kinematics of the growing mixture}\label{sec:kinematics}
We consider the  so--called \emph{intermingled mixture constraint}~\cite{AmbrosiPreziosi},
stating that all the solid matrix constituents
experience the same overall motion. This hypothesis
requires the displacement and velocity vectors of each constituent
to be equal to those of the solid matrix. Then, we denote by:
\begin{subequations}\label{eq:kinematics}
\begin{align}
&
\mathbf{u}_{\s}=\mathbf{u}_{\s}(\mathbf{x},t) & \label{eq:displacement} \\
& \mathbf{v}_{\s}=\mathbf{v}_{\s}(\mathbf{x},t)=
\dfrac{\partial}{\partial t}\mathbf{u}_{\s}(\mathbf{x},t)
& \label{eq:velocity}
\end{align}
the displacement and velocity at the time level $t$
of any point $\mathbf{x}$ of the solid component of the biomass,
and by
\begin{align}
&
\boldsymbol{\varepsilon}_{\s}(\mathbf{x},t)
= \frac{1}{2} (\nabla \mathbf{u}_{\s}(\mathbf{x},t) +
(\nabla \mathbf{u}_{\s}(\mathbf{x},t))^T) & \label{eq:strain}
\end{align}
the associated infinitesimal deformation of the biomass volume
surrounding the point $\mathbf{x}$ at time $t$.
The intermingled mixture constraint
yields also the following relation
\beq\label{eq:eps_eta_eps_s}
\boldsymbol{\varepsilon}_{\eta}= \boldsymbol{\varepsilon}_{\s} \qquad  \eta=\cells,\ECM.
\eeq
The growth process of each
mixture component (cellular growth and ECM secretion) is taken into account by
introducing the following decomposition~\cite{Klisch}
\beq\label{epsilonbzeta}
\boldsymbol{\varepsilon}_{\eta}=\boldsymbol{\varepsilon}_{\eta}^g+
\boldsymbol{\varepsilon}_{\eta}^e, \qquad \eta=\cells,\ECM,
\eeq
where $\boldsymbol{\varepsilon}^g_{\eta}$ is the infinitesimal growth tensor
associated with each solid constituent of the biomass,
accounting for the amount
and the spatial orientation of the newly deposited mass,
and $\boldsymbol{\varepsilon}^e_{\eta}$ is the elastic accommodation tensor
necessary to reinforce at each time level
the continuity of the whole solid upon growth. Finally, we denote by
\beq\label{relative_velocity}
\mathbf{w}=\mathbf{v}_{\fl}-\mathbf{v}_{\s}
\eeq
\end{subequations}
the relative (diffusive) velocity~\cite{Tosin,Sengers} of the fluid
phase with respect to the solid phase in the biomass, $\mathbf{v}_{\fl}$
being the velocity of the interstitial fluid.
Phenomenological laws for $\boldsymbol{\varepsilon}^g_{\eta}$,  $\boldsymbol{\varepsilon}^e_{\eta}$
and $\mathbf{w}$ are discussed in Sect.~\ref{sec:bio-mechanical_models}.
For notational brevity, from now on,
we simply write $\mathbf{u}$ and $\boldsymbol{\varepsilon}$ instead of $\mathbf{u}_{\s}$ and $\boldsymbol{\varepsilon}_{\s}$, respectively.

\section{Balance laws for the deformable growing biomass}\label{sec:balance_laws}
In this section we illustrate the set of conservation laws
that constitute our proposed
mathematical picture of the mechanobiological
processes regulating biomass tissue growth.

\subsection{Mass balance for the growing
biomass}\label{sec:mass_balance}
The mass balance equation for the growing mixture
is expressed by the following coupled system of PDEs
in conservation form to be solved in $\mathcal{Q}_{T_{end}}$:
\begin{subequations}\label{eq:mixture_mass_balance}
\begin{align}
& \frac{\partial \boldsymbol{\phi}}{\partial t}+
\textbf{div}\,\mathbf{J}_{\boldsymbol{\phi}} =
\mathbf{Q}(\boldsymbol{\phi},c,\mathbf{T})
& \label{eq:continuity} \\
& \mathbf{J}_{\boldsymbol{\phi}} =
\left[\phi_{\fl} \mathbf{v}_{\fl},\phi_\n \mathbf{v}_{\s},\phi_\v \mathbf{v}_{\s},\phi_\q \mathbf{v}_{\s},\phi_{\ECM} \mathbf{v}_{\s}\right]^T
& \label{eq:flux} \\
& \mathbf{Q} = \left[Q_{\fl}, Q_{\n}, Q_\v, Q_\q,
Q_{\ECM}\right]^T &\label{eq:prod_term}
\end{align}
where $\mathbf{J}_{\boldsymbol{\phi}} \in
\mathbb{R}^{5 \times d}$ is the flux matrix and
$\mathbf{Q}$
is the net production rate, whose  mathematical form
is described in detail in Sect.~\ref{sec:production_terms}.
Due to Assumption~\ref{ass:closedsystem},
the term~$\mathbf{Q}$ satisfies the following constraint
\beq\label{eq:sum_Qi}
\sum_{\zeta=\s,\fl}Q_\zeta=0.
\eeq
Eq.~\eqref{eq:flux} represents a phenomenological
description of the flux density of each species under the effect
of convective transport due to the fluid and solid velocity,
respectively.
It is worth noting that in cartilage tissue growth,
cells do not typically exhibit a significant diffusive motion,
rather, they need a solid support for 
\rs{surviving and for} developing their
functional activities (property of anchorage-dependence~\cite{Nikolaev}).
For this reason in the present work we neglect the contribution to the flux density
due the the diffusive transport, unlike in other applications where
this term plays a significant role~\cite{Moreo2,Murray,Maini2002}.
\end{subequations}

\subsection{Momentum balance for the growing mixture}\label{sec:momentum}
Under the assumption of negligible inertial terms and absence of
body forces and volumetric fluid mass sources/sinks,
the linear momentum balance equation for the solid and fluid
phases of the growing mixture is expressed by the following PDEs
in conservation form to be solved in $\mathcal{Q}_{T_{end}}$:
\begin{subequations}\label{eq:biomass_linear_momentum}
\begin{align}
& \textbf{div}{}\mathbf{T}_{\zeta}
(\mathbf{u}, \, p, \, \boldsymbol{\phi}) + \boldsymbol{\pi}_\zeta=
\mathbf{0} & \qquad \zeta = \s, \fl
\label{eq:mech_balance2} \\[2mm]
& \mathbf{T}_{\s}(\mathbf{u}, \, p, \, \boldsymbol{\phi}) =
\sum_{\eta=\cells,\ECM}\phi_{\eta}
\mathbf{T}_{\eta}(\mathbf{u}, \, p, \, \boldsymbol{\phi})
&  \label{eq:stress_solid2} \\
& \mathbf{T}_{\eta}(\mathbf{u}, \,p, \, \boldsymbol{\phi}) =
\boldsymbol{\sigma}_{\eta}(\mathbf{u}, \, \boldsymbol{\phi})
- p \mathbf{I} & \qquad \eta=\cells,\ECM
\label{eq:stress_solid3} \\[2mm]
& \mathbf{T}_{\fl}(\mathbf{u}, \, p, \, \boldsymbol{\phi}) =
- \phi_{\fl} \, p \mathbf{I},
& \label{eq:stress_fluid2}
\end{align}
where $\boldsymbol{\sigma}_\eta$ is the effective stress tensor of the
component $\eta$ of the solid phase of the mixture,
$p=p(\mathbf{x},t)$ is the pressure exerted by the
fluid phase and $\mathbf{I}$ is the identity
tensor. The isotropic stress $-p  \mathbf{I}$ accounts for the
coupling, typical of poroelasticity, between the
flow of the fluid and the deformation of the solid matrix, and in particular describes the contribution to the stress due to the fluid pressure within the structure.

The quantities $\mathbf{T}_{\zeta}$, $\zeta = \s, \fl$,
are the total stress tensors of the solid and fluid phases, while
$\boldsymbol{\pi}_\zeta$ are the interphase forces~\cite{Lemon2006}. As usual,
we neglect the effective stress tensor of the fluid, meaning
that we assume that the internal fluid viscosity is negligible
compared with the friction between the fluid and the solid matrix~\cite{Barry,frijns,Tosin}.
For the mathematical characterization of the forces $\boldsymbol{\pi}_{\zeta}$ we
refer to~\cite{Lemon2006} and~\cite{Tosin}. We observe that,
for all $t \in (0, T)$ and at all $\mathbf{x}\in \Omega$,
it holds
\begin{align}
& \boldsymbol{\pi}_{\s}(\mathbf{x},t)
+ \boldsymbol{\pi}_{\fl}(\mathbf{x},t) = \mathbf{0}.
& \label{eq:interphase_force_zero}
\end{align}
\end{subequations}

\subsection{Total mass and momentum balance for the growing biomass}
\label{sec:total_mass_momentum_balance}
The mass balance equation for the whole growing mixture
is obtained by summing each component in system~\eqref{eq:continuity} and
using Assumptions~\ref{ass:fullysaturation}
and~\ref{ass:closedsystem}, to get:
\begin{subequations}\label{total_equations}
\begin{align}
& \mathrm{div}\,\mathbf{v} = 0 & \label{eq:continuity_global} \\
& \mathbf{v} = \phi_{\fl} \mathbf{v}_{\fl}
+ \phi_{\s} \mathbf{v}_{\s}.
& \label{eq:composite_velocity}
\end{align}
The vector $\mathbf{v}$ is the composite velocity of the mixture
(cf. Eq. (2.4) of~\cite{Tosin}) and Eq.~\eqref{eq:continuity_global}
expresses the conservation of the total mass of the growing tissue.
A simple manipulation allows us to write
Eq.~\eqref{eq:continuity_global} as
\begin{align}
& \Frac{\partial}{\partial t}\textrm{div}{} \mathbf{u} +
\mathrm{div}(\phi_{\fl} \mathbf{w}) = 0.
& \label{eq:continuity_2}
\end{align}
In a similar manner,
summing equations~\eqref{eq:biomass_linear_momentum}
and using~\eqref{eq:interphase_force_zero}, we get
\begin{align}
& \mathbf{div}{}\mathbf{T}(\mathbf{u}, \, p,
\, \boldsymbol{\phi})
= \mathbf{0} & \label{eq:momentum_balance} \\
& \mathbf{T}(\mathbf{u}, p, \, \boldsymbol{\phi})
= \sum_{\eta= \cells, \textrm{ECM}} \phi_{\eta}\boldsymbol{\sigma}_{\eta}
(\mathbf{u}, \, \boldsymbol{\phi})- p \mathbf{I}. &
. \label{eq:total_stress}
\end{align}
\end{subequations}
The quantity $\mathbf{T}$ is the total stress in the mixture and
Eq.~\eqref{eq:momentum_balance} expresses
the conservation of the total momentum of the growing tissue.

\subsection{Mass balance for nutrient concentration}
\label{sec:mass_balance_oxygen}

The mass balance system~\eqref{eq:mixture_mass_balance}
for the solid and fluid phases of the growing mixture
is accompanied by a corresponding
continuity equation for the nutrient concentration (oxygen)
$c=c(\mathbf{x},t)$ that is transported
throughout the growing on mixture by the interstitial fluid.
This continuity equation is expressed by
the following PDE in conservation form to be solved
in $\mathcal{Q}_{T_{end}}$:
\begin{subequations}\label{oxygen_mass_balance}
\begin{align}
& \frac{\partial c}{\partial t}+
\textrm{div}\,\mathbf{J}_\mathrm{c} = Q_\mathrm{c}(\boldsymbol{\phi},c)
& \label{eq:continuity_oxygen} \\
& \mathbf{J}_\mathrm{c} = \mathbf{v}_{\fl} c -D_\mathrm{c} \nabla c, &
\label{eq:flux_oxygen}
\end{align}
\rs{the interstitial fluid velocity $\mathbf{v}_{\fl}$ being
computed using~\eqref{relative_velocity} as
\begin{align}
& \mathbf{v}_{\fl} = \mathbf{w} +
\mathbf{v}_{\s} = \mathbf{w} +
\dfrac{\partial\mathbf{u}}{\partial t}. & \label{eq:v_fl}
\end{align}}
The mathematical description of the oxygen diffusion coefficient $D_\mathrm{c}$
adopted in this article is the so-called Maxwell model~\cite{Wood2002},
that allows to account, in a volume-averaged sense, for the
microscopic composition of the biomass. More precisely, we
introduce the effective diffusion coefficient
\be
D_\mathrm{c}:=D_{\mathrm{c},\fl}\dfrac{3k-2\phi_\fl(k-1)}{3+\phi_\fl(k-1)},
\qquad k:=K_\mathrm{eq}\dfrac{D_{\mathrm{c},\s}}{D_{\mathrm{c},\fl}}
\ee
where $D_{\mathrm{c},\fl}$ and $D_{\mathrm{c},\s}$ represent the nutrient diffusivity in the
fluid and solid phase, respectively, while $K_\mathrm{eq}$
is the coefficient regulating local mass equilibrium between
nutrient concentration in the solid and fluid phases
(see~\cite{Wood2002} for a detailed discussion).

The time rate of oxygen consumed by the
cellular populations is modeled by a generalized form
of the Michaelis-Menten kinetics
\begin{align}
& Q_c(\boldsymbol{\phi},c) =
- (R_\n \phi_\n + R_\v \phi_\v + R_\q \phi_\q)
\Frac{c}{c+K_{1/2}} & \label{eq:MM}
\end{align}
where $R_\eta$, $\eta=\n,\v,\q$, is the nutrient consumption rate
of the cellular population $\phi_\eta$
and $K_{1/2}$
is the half saturation constant.
We refer to~\cite{Sacco} and the literature cited therein
for a similar treatment of the oxygen consumption term in the
framework of a multi-phase growing mixture.
\end{subequations}

\section{Mass exchange pathways}
\label{sec:production_terms}
The production terms $Q_{\eta}$
introduced in Eq.~\eqref{eq:prod_term}
mathematically describe the mechanisms of
addition and/or removal of mass for each species
constituting the biomass growing mixture.

\begin{figure}[h!]
\centering
\includegraphics[width=0.8\textwidth]{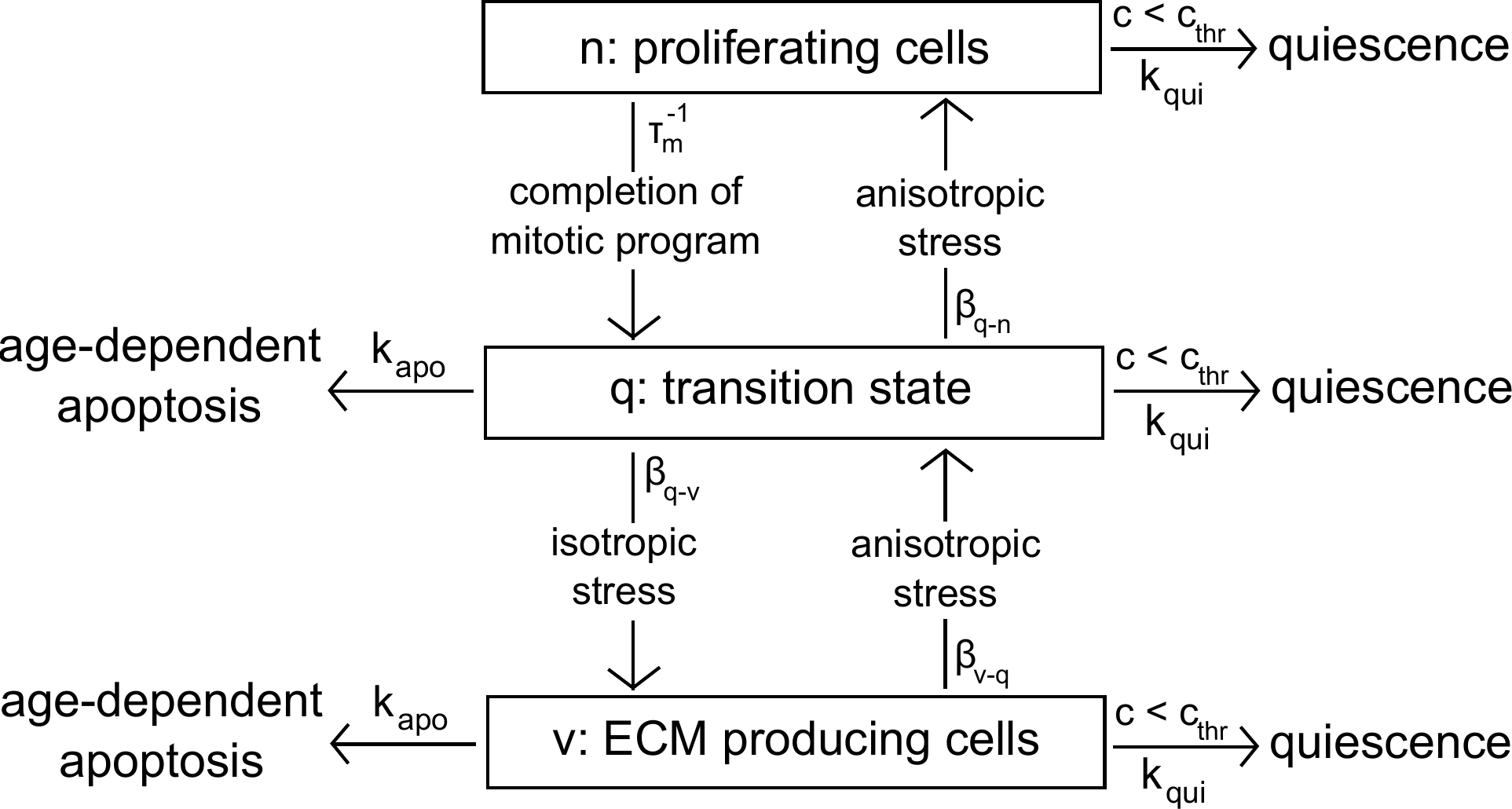}
\caption{Conceptual scheme of exchange pathways
among cellular populations
(generalization of Figure 5.3 in~\cite{SengersPhD}).}
\label{fig:schema_n_v_q}
\end{figure}

The exchange pathways between the
different functional cellular pools are supposed
to be mediated by the local stress state, the local nutrient
concentration and by natural decay times.
Transitions among cellular compartments
are modeled by the time rate (probability) of exchange $\beta_{\alpha-\gamma}$
between states $\alpha$ and $\gamma$, $\alpha,\gamma=\n,\v,\q$, while natural decay
is modeled by decay rate constants $k_{\text{apo}},k_{\text{qui}}$.
The scheme of Fig.~\ref{fig:schema_n_v_q} illustrates the
various transitions among constituents, that may occur according
to the following cases:
\begin{subequations}\label{eq:ABtransitions}
\begin{itemize}
\item[-] stress-mediated exchange pathways: cells in state~$\q$
may switch into proliferative state or ECM secreting with a probability
related to the stress state they experience.
An anisotropic stress state enhances transition towards the proliferative state,
while an isotropic stress state enhances transition towards the ECM secreting state.
We quantify this concept as follows. Let $H(z)$ be the Heaviside function such that $H(z)=0$ for
$z < 0$ and $H(z)=1$ for $z \geq 0$, $z$ being a real variable. Then, the stress-state dependent
transition rate is represented by
\begin{align}
& \beta_\text{A$\rightarrow$B} H(r - \overline{r}),
\label{eq:trans_heavyside}
\end{align}
$r$ being a yet unspecified indicator of the isotropy/anisotropy of the local stress
state and $\overline{r}$ being a threshold value.
\rs{An explicit characterization of the local stress indicator $r$
will be provided in Part II of the present article.}
In our modeling description,
if $r < \overline{r}$ the local state of stress can be considered
as isotropic, while if $r \geq \overline{r}$ the local state of stress
can be considered as anisotropic.
\item[-] mitosis-mediated exchange pathway:  we suppose cell
proliferation to be expressed by the following law
\begin{align}
& \phi_\n(1-(\phi_\n+\phi_\v+ \phi_\q+\phi_{\ECM}))
\dfrac{c}{K_\text{sat}+c} k_\text{g} =
\phi_\n \phi_{\fl} \dfrac{c}{K_\text{sat}+c} k_\text{g}.
\label{eq:cell_proliferation}
\end{align}
Relation~\eqref{eq:cell_proliferation} is a
phenomenological law given by the product of two terms:
the first term, given by $\phi_\n \phi_{\fl}$,
keeps into account contact inhibition effects, while
the term $\displaystyle \frac{c}{K_\text{sat} +c}k_\text{g}$
is a nutrient--dependent modulation (Monod law~\cite{Contois}),
$K_\text{sat}$ being the half-saturation constant and
$k_\text{g}$ the maximum growth rate, respectively;
\item[-] decay pathways: all cellular compartments may
evolve into quiescent (absence of cell activity due to an
insufficient oxygen
intake~\cite{Coller,SengersPhD}) or apoptotic phases
(cellular death). Quiescence occurs if the nutrient concentration~$c$ falls below a critical level~$c_{thr}$, whereas the apoptotic phase
is related to age dependent cell death~\cite{Tew2001}.
The time rates of change between state $\alpha$ ($\alpha=\n,\v,\q$) and the inactive states (quiescence or apoptosis) are
$k_\qui$ and $k_\apo$, respectively.
Moreover, the transition rates
${\n\leftrightarrow \q}$
are regulated by the mitotic characteristic time rate
$1/\tau_{\rm m}$ and take the form
$\mp \Frac{\phi_n}{\tau_{\rm m}}$, where the minus sign is used in the $\n\rightarrow \q$ transition, the plus sign in the
$\q\rightarrow \n$ transition.
\end{itemize}
\end{subequations}
For notational brevity we set $H_r:=H(r-\bar{r})$ and
$H_c:=H(c-c_\text{thr})$.
According to the exchange laws illustrated above,
the production terms associated with cell populations are defined as:
\begin{subequations}\label{eq:production_terms}
\begin{align}
&Q_\n =-\dfrac{\phi_\n}{\tau_\text{m}}+
\phi_\q\beta_{\q \rightarrow \n} H_r +\phi_\n \phi_{\fl}
\dfrac{c}{K_\text{sat}+c} k_\text{g} -k_\qui\phi_\n(1-H_c)
&\label{eq:n_production}\\[3mm]
&Q_\v =-\phi_\v\beta_{\v\rightarrow \q}H_r+
\phi_\q \beta_{\q\rightarrow \v}
\left(1-H_r \right)-k_\qui\phi_\v(1-H_c) -
k_\text{apo}\phi_\v&\label{eq:v_production}\\[3mm]
&Q_\q=\dfrac{\phi_\n}{\tau_\text{m}}-
\phi_\q\beta_{\q\rightarrow \n}H_r+\phi_\v
\beta_{\v\rightarrow \q}H_r-\phi_\q
\beta_{\q\rightarrow \v}\left(1-H_r\right)&\notag\\
&-k_\qui\phi_\q\left(1-H_c\right)-
k_\text{apo}\phi_\q.&\label{eq:q_production}
\end{align}

As for the definition of the production term for ECM mass density,
we describe the biosynthesis of the different ECM components,
focusing particularly on glycosaminoglycan (GAG) and collagen.
We consider here
GAG as the main marker for ECM accumulation~\cite{Nikolaev} and we
assume a simple proportionality law between the total amount
of ECM and the secretion rate of GAG.
The constant of proportionality $E>1$ accounts for the
heterogeneous composition
of cartilagineous ECM (water for 70-80\% of its wet weight,
collagen fibrils for 10-15\% and GAG for 5\%)~\cite{Causin}.
\rs{According to our model, ECM synthesis attains its maximum value
when no extracellular matrix is present because more space is available
for matrix production. Then, as soon as sythesized matrix accumulates 
at each point of the biomass construct, the available space diminuishes
until $\phi_\ECM$ reaches a maximum value $\phi_\text{ECM,max}$ and 
matrix secretion ceases. Therefore $Q_{\ECM}$ takes the following form:}
\beq\label{eq:GAG_production}
\begin{split}
Q_{\ECM} &= \dfrac{\phi_\v}{V_{\cell}} \, c \,E\, k_\text{GAG} \max\Big[0,1-\dfrac{\phi_\text{ECM}}{\phi_\text{ECM,max}}\Big]-
k_\text{deg}\phi_\text{ECM}\\
&= \begin{cases}
\dfrac{\phi_\v}{V_{\cell}} \, c \,E\, k_\text{GAG}\Big(1-\dfrac{\phi_\text{ECM}}{\phi_\text{ECM,max}}\Big)
-k_\text{deg} \phi_\text{ECM}&
\text{ if }\phi_\text{ECM} \leq \phi_\text{ECMmax}\\
-k_\text{deg}\phi_\text{ECM}&\text{ if }\phi_\text{ECM}>
\phi_\text{ECM,max}\\
\end{cases},
\end{split}
\eeq
where $V_{\cell}$ is the volume of a single cell involved in the
ECM secretion process while
$k_\text{deg}$ is the ECM degradation rate~\cite{Please}.
In the description of GAG secretion,
we are assuming that at the initial time level, biomass is constituted by a uniform layer of cells and matrix (see~\cite{Raimondi2008} for a similar ap-
proach). This corresponds to neglecting
the very initial phase where the seeded cells proliferate and
``pave'' the scaffold wall, and is consistent with the mathematical fact that
a continuum-based approach does not enable to reproduce the subcellular mechanisms that regulate the early mitotic process.
These latter processes should be more properly described by treating
seeded cells as individual units that behave according to cellular automata schemes~\cite{Chung,Cheng,Galbusera2007,Galbusera2008}.

To conclude the mathematical description of mass exchange terms,
we consider the extracellular fluid production $Q_{\fl}$. This latter
quantity is defined in such
a way to satisfy Assumption~\ref{ass:closedsystem} and,
consistently, relation~\eqref{eq:sum_Qi}.
To this purpose, $Q_{\fl}$ is modeled as
\beq\label{eq:fluid_production}
Q_{\fl}=-\sum_{\eta=\cells, \ECM} Q_{\eta}.
\eeq
From a biophysical point of view this is equivalent to assuming that mass exchanges occur only among cells/ECM and fluid, meaning that dead cells and degrading ECM are deteriorated into extracellular fluid, and conversely that
the latter is \rs{"consumed"} whenever cells duplicate or secrete ECM~\cite{Tosin2}.
From a computational point of view, relation~\eqref{eq:fluid_production}
allows us to eliminate the dependent variable
$\phi_{\fl}$ and the corresponding mass balance equation
from system~\eqref{eq:continuity}
as done in~\cite{Tosin2}, Sect.~2.2,
in such a way that the fluid volume fraction can be computed by
simple post-processing as
\beq\label{eq:phi_fl_evaluation}
\phi_{\fl} = 1 - \sum_{\eta=\cells, \ECM} \phi_{\eta}.
\eeq
\end{subequations}

\section{Bio-mechanical models for the deformable growing biomass}
\label{sec:bio-mechanical_models}
In this section we provide a mathematical description
of the mechanobiological processes involving cell populations and
ECM secretion. To this purpose, we introduce suitable bio-mechanical models
for the growth tensors in the decomposition~\eqref{epsilonbzeta}
by extending the theory developed in~\cite{Klisch0} and~\cite{Klisch}.
\begin{figure}[h!]
\centering
\includegraphics[width=0.8\textwidth]{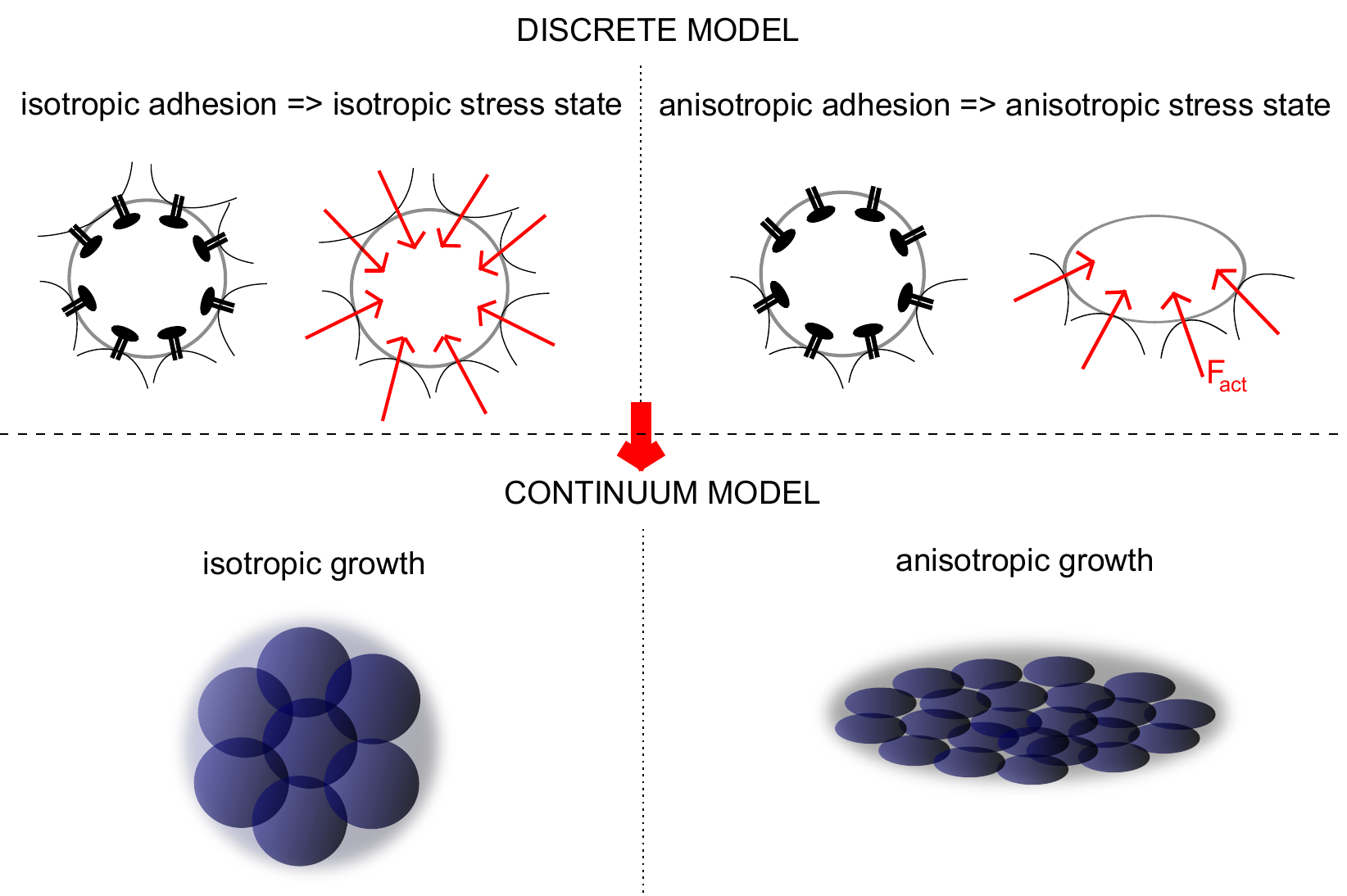}
\caption{\rs{Cellular level (top):  pictorial representation of the isotropic/anisotropic adherence condition. 
Continuum level (bottom): isotropic/anisotropic biomass growth.}}
\label{fig:iso_aniso_growth_2}
\end{figure}

\subsection{Growth laws}\label{sec:growth_laws}
We propose the following definitions of the growth tensors:
\begin{subequations}\label{eq:growth_tensors}
\begin{align}
& \boldsymbol{\varepsilon}^g_{\vartheta}
(\textbf{x},t; \boldsymbol{\phi}) =
g_\theta(\mathbf{x},t; \boldsymbol{\phi})\mathbf{I} &
\qquad \vartheta = \v, \q, \ECM \label{eq:growth_1}\\
& \boldsymbol{\varepsilon}^g_\n (\textbf{x},t; \boldsymbol{\phi}) =
g_\n(\mathbf{x},t; \boldsymbol{\phi})\mathbf{d}^{\text{pol}}
(\textbf{x},t)\otimes\mathbf{d}^{\text{pol}}
(\textbf{x},t) \label{eq:growth_2}
\end{align}
where the symbol $\otimes$ represents the tensor dyadic product
\rs{and $g_\theta$, $g_\n$ are growth coefficients for which a model
equation is provided below.}
Eqns.~\eqref{eq:growth_tensors} state that the mass increment of each
mixture solid constituent
is isotropically deposited for the $\v$, $\q$ and $\ECM$
components, while is accumulated along a specific polarization direction, identified by the unit vector $\mathbf{d}^\text{pol}$,
for proliferating cells.
Let us now address some biophysical motivations that support
our choice of the growth laws~\eqref{eq:growth_tensors}.
Firstly, according to the concept of ``force isotropy'' on the cell
introduced in~\cite{Nava2012,Nava2014}, cells that occupy the bio-synthesizing compartment ($\v$ compartment) experience
an isotropic adherence condition and consequently
tend to assume a spherical shape (see Fig.~\ref{fig:iso_aniso_growth_2}, left)
whereas cells that live in the proliferating compartment
($\n$ compartment) are subjected to an anisotropic adhesion state and tend to elongate (see Fig.~\ref{fig:iso_aniso_growth_2}, right).
Secondly, according to the infinitesimal deformation growth theory developed in~\cite{Klisch0}, the deformation of an infinitesimal sphere
of biomass growing into an ellipsoid
can be reasonably described by an anisotropic growth tensor, while the deformation of an infinitesimal sphere growing into a larger sphere
can be characterized by a isotropic growth tensor.
For the sake of simplicity, the infinitesimal growth tensor for the species
$\q$ and $\ECM$ are supposed to be isotropic.

The definition of $\mathbf{d}^\text{pol}$ in Eq.~\eqref{eq:growth_2} and the law for its time
evolution is a delicate issue.
In~\cite{Moreo3D}, $\mathbf{d}^\text{pol}$ is
characterized according to the dynamics of the evolution of the cell cytoskeleton,
which reorganizes itself according to its mechano--sensing mechanisms.
A simplified version of the model proposed in~\cite{Moreo3D},
and also adopted in the present work, is represented
by the choice
\beq\label{eq:d_pol}
\mathbf{d}^\text{pol}(\mathbf{x},t)= \mathbf{d}_{\varepsilon}
(\mathbf{x},t) \qquad \forall \mathbf{x} \in \Omega, \quad \forall t>0
\eeq
where $\mathbf{d}_{\varepsilon}$ is the normalized eigenvector of
the infinitesimal strain tensor $\boldsymbol{\varepsilon}_{\s}$,
associated with the eigenvalue of largest module,
which physically corresponds to the maximum principal dilatation of
the biomass around such a point~\cite{sokolnikoff1956}.

The coefficients $g_{\eta}(\mathbf{x},t; \boldsymbol{\phi})$,
$\eta = \n, \v, \q, \ECM$, give a measure of the amount of
mass of the cellular population of type $\eta$ deposited
at time $t$ at point $\mathbf{x}$. To determine these
quantities, we proceed as in~\cite{Klisch0} and~\cite{Klisch}
and require the following growth continuity initial value problem
to be satisfied for all $\mathbf{x} \in \Omega$:
\begin{align}
& \dfrac{\partial}{\partial t} \text{Tr}{}\boldsymbol{\varepsilon}^{\text{g}}_{\eta}
(\mathbf{x},t; \boldsymbol{\phi}) =
c_{R,\eta}(\mathbf{x},t; \boldsymbol{\phi}) & \qquad t \in (0, T_{end}]
 \quad\eta=\cells,\ECM, \label{growth_continuity1} \\
& \text{Tr}{}\boldsymbol{\varepsilon}^{\text{g}}_{\eta}
(\mathbf{x},0; \boldsymbol{\phi}) = 0. & \label{initial_condition_growth}
\end{align}

The quantity $c_{R,\eta}(\mathbf{x},t; \boldsymbol{\phi})$
represents the amount of mass of the cellular population $\eta$
deposited at time $t$ at point $\mathbf{x}$
per unit time and per unit reference mass.
According to the general indications illustrated in
Sect. 2.2.4 of~\cite{Klisch}, the growth laws are
phenomenological equations that indirectly describe
chemical processes responsible for growth and can be typically
expressed as "synthesis" rate minus a "degradation" rate, that
may include a mass conversion rate from one constituent of the mixture
to another. Also, the constants that appear in a specific
growth law may depend parameterically \rs{on} biological factors such as,
for example, the level of a specific growth factor.
Thus, based on the description carried out in
Sect.~\ref{sec:production_terms}, we set
\begin{align}
c_{R,\eta}(\mathbf{x},t; \boldsymbol{\phi}):=
Q_\eta(\mathbf{x},t; \boldsymbol{\phi}) & \qquad \eta=\cells,\ECM,
\label{eq:definition_of_cR}
\end{align}
in such a way that the initial value problems that furnish the
characterization of the growth coefficients become:
\begin{align}
& \dfrac{\partial}{\partial t}g_\theta(\mathbf{x},t; \boldsymbol{\phi})
= \Frac{1}{3} Q_\theta(\mathbf{x},t; \boldsymbol{\phi})
& \qquad t \in (0, T_{end}]
 \quad\theta=\v,\q, \ECM, \label{growth_continuity_vqE} \\
& g_\theta(\mathbf{x},t; \boldsymbol{\phi}) = 0 & \label{initial_condition_growth_vqE}
\end{align}
and:
\begin{align}
& \dfrac{\partial}{\partial t}g_\n(\mathbf{x},t; \boldsymbol{\phi})
= Q_\n(\mathbf{x},t; \boldsymbol{\phi})
& \qquad t \in (0, T_{end}] \label{growth_continuity_n} \\
& g_\n(\mathbf{x},t; \boldsymbol{\phi}) = 0 &
\label{initial_condition_growth_n}
\end{align}
having used the identities
$\text{Tr}(\mathbf{I}) = 3$ and
$\text{Tr}(\mathbf{d}^\text{pol}\otimes\mathbf{d}^\text{pol})=1$.
\end{subequations}

\subsection{Constitutive equations for the mechanical and fluid
subsystems}\label{sec:Constitutive_equations}

We assume that cells and ECM behave like linear elastic solids,
so that the effective stress tensors associated with
the solid components of the biomass are defined as
\begin{subequations}\label{eq:constitutive_laws}
\begin{align}
& \boldsymbol{\sigma}_{\eta}(\mathbf{u}, \, \phi_{\eta}) =
2 \mu_\eta \boldsymbol{\varepsilon}_{\eta}^{\e}(\mathbf{u}, \, \phi_{\eta})
+ \lambda_\eta \text{Tr}{}
\boldsymbol{\varepsilon}_{\eta}^\e(\mathbf{u}, \, \phi_{\eta})
\mathbf{I} & \qquad \eta=\n,\v,\q, \ECM. \label{eq:stress_cells}
\end{align}
Recalling relation~\eqref{epsilonbzeta}, Eq.\eqref{eq:stress_cells} can be written as
\begin{align}
\boldsymbol{\sigma}_{\eta}(\mathbf{u}, \, \phi_{\eta}) =
2 \mu_\eta \left(\boldsymbol{\varepsilon}(\mathbf{u})
-\boldsymbol{\varepsilon}_{\eta}^g(\phi_{\eta}) \right)
+ \lambda_\eta \text{Tr}{}
\left(\boldsymbol{\varepsilon}(\mathbf{u})
-\boldsymbol{\varepsilon}_{\eta}^g(\phi_{\eta})\right)\mathbf{I}
\end{align}
where $\lambda_{\eta}$ and $\mu_\eta$ are the Lam\'e parameters
of each component of the solid phase, $\eta=\n, \q, \v, \ECM$,
and $\boldsymbol{\varepsilon}_{\eta}^g(\phi_{\eta})$ are the
growth strain tensors introduced in~\eqref{eq:growth_tensors}.
More sophisticated constitutive models might
be adopted~\cite{Tosin,Moreo1,DiMilla1991,Araujo}, but their use
is beyond the scope of the present work which is mainly devoted
to proposing a computationally feasible
mechanobiological model of {\em in vitro} cartilage tissue growth.

We assume the relative velocity
in Eq.~\eqref{eq:continuity_2} to be expressed by the Darcy
law (see, e.g.,~\cite{ByrnePreziosi2003} and references cited therein)
\beq\label{constitutive_w}
\phi_{\fl}\mathbf{w}=-\mathbf{K}\nabla p
\eeq
where $\mathbf{K}=\mathbf{K}(\phi_\fl)$ is the isotropic permeability tensor
defined as in~\cite{Sengers}
\beq\label{permeability_tensor}
\mathbf{K}(\phi_{\fl})=\frac{\phi_{\fl}^2}{C_\text{F}} \mathbf{I},
\eeq
the quantity $C_\text{F}$ being a friction coefficient. To provide a
physically consistent characterization of
$C_\text{F}$ we apply the
classic Stokes theory for viscous drag to the biomass mixture
and obtain
\begin{align}
C_\text{F} = C_\text{F,cell}
\phi_{\s} = \frac{6 \pi \mu_{\fl}}{A_\text{cell}} (1- \phi_{\fl})
= \frac{3 \mu_{\fl}}{2 R_\text{cell}^2}  (1- \phi_{\fl}) \label{eq:C_F}
\end{align}
$R_\text{cell}$ and $\mu_{\fl}$ being cell radius and interstitial fluid
dynamic viscosity, respectively. Replacing~\eqref{eq:C_F}
into~\eqref{permeability_tensor} we can express the biomass permeability
tensor as
\beq\label{permeability_tensor_2}
\mathbf{K}(\phi_{\fl})= K_\text{ref} \frac{\phi_{\fl}^2}{1- \phi_{\fl}}\mathbf{I}
\eeq
where
\beq\label{permeability_ref}
K_\text{ref} = \frac{2}{3} \frac{R_{cell}^2}{\mu_{\fl}}
\eeq
is a reference value of biomass permeability.
\end{subequations}

\section{\rs{Conclusions and research perspectives}}
\label{sec:conclusions_part_1}

In the present article, which is the first part of a series of two
distinct but correlated contributions, we propose a novel mathematical
formulation based on the continuum assumption to describe the biomechanical
sensitivity \rs{of articular chondrocytes.}
The natural application of our model is \rs{Tissue Engineering}, a continuously growing
discipline within the wider area of \rs{Regenerative Medicine,} in which
the control of cell response to multi-factorial stimuli is of 
utmost importance
to obtain products suitable to clinical practice. However, it is worth 
noting that the proposed scheme may be
used as well to describe more general settings in Cellular Biology, for example,
the \rs{expansion} of staminal cells.

\null

The principal novelty of our contribution is the development of a model
based on the use of Partial Differential Equations (PDEs) that incorporates
the concept of ``force isotropy'' on the cell within the general and well established
framework of poroelastic theory of mixtures and of cell population models.
Specifically, the model translates into a simplified mathematical formalism,
based on the use of suitably parametrized Heaviside functions, how
the induced cytoskeletal tensional states trigger signalling transduction
cascades regulating functional cell behavior, for example, the 
\rs{traslocation} of specific
transcription factors in the nucleus. According to the concept of force isotropy,
it turns out that if cell adhesion-mediated
traction forces have approximately the same strength over the cell surface,
then the cell nucleus tends to maintain a roundish morphology, while
if cell adhesion-mediated traction forces have different
magnitudes at varying spatial orientations on the cell surface,
then the cell nucleus tends to elongate. In the first case, the cell
tensile condition is defined as ``isotropic cytoskeletal tension'' whereas in the second
case the cell tensile condition is defined as ``anisotropic cytoskeletal tension''.

\null
Having defined the cytoskeletal stress characterization at the single cellular level,
the next step of our approach is to build upon the concept of continuum-based approach to
extend the above described description to the local stress tensor associated with
the biological construct to mathematically represent the isotropic/anisotropic cell adhesion state.
To this purpose, we generalize in a natural manner the previous definitions prescribing that
if the anisotropic part of the local stress tensor is lower than a fixed threshold
then the local stress state of the system is isotropic otherwise the local stress state of the system is anisotropic.

\null

The final step of our model construction is to incorporate the above illustrated
mechanobiological scheme within the setting of the theory of poroelasticity of a mixture
composed by a solid and a multi-component fluid phases. The mixture represents the cellular
construct in which several different cellular populations are well-mixed and oxygen delivery
and consumption is taken into account to regulate in a dynamical manner the progressive
fate of the evolving (macroscopic) tissue.
The overall mathematical formulation consists of a system of conservation laws
(mass and linear momentum) for
the phases and components of the mixture that includes stress state and
oxygen tension as main determinants of cellular culture evolution.

\null
The high level of complexity of our model requires a severe effort for: 
\begin{enumerate}
\item its mathematical analysis (well posedness and linear stability); 
\item its numerical simulation (in a truly 3D geometrical configuration). 
\end{enumerate}

\rs{As far as issue 1. is concerned,
we shall illustrate in a subsequent paper}
a stability analysis looking for homogeneous stable
steady states of the dynamical system that 
describes the conservation of mass of the solid mixture components.
Such a study will allow us to characterize the admissible range of values of
model constitutive parameters that ensures the 
\rs{biophysical} consistency of the
proposed mathematical representation, in the same spirit as in~\cite{Moreo2} and~\cite{Maini2002}.

As far as issue 2. is concerned, we refer the reader 
to the second part of the present article where 
a thorough investigation of the PDE system is critically performed in a simplified 1D setting.

\rs{Further} research effort will be devoted to extend the present formulation by including the following advanced mechanobiological aspects:
\begin{itemize}
\item in the case of basic Cellular Biology, the contribution of the active
forces exerted by the cell to test the mechanical properties of the surrounding
environment (cf.~\cite{Moreo1} and~\cite{Tosin2});
\item in the case of TE applications,
the coupling between the growing construct with the surrounding 
\rs{perfused culture} fluid
(cf.~\cite{Lemon2007} and~\cite{Sacco,Causin}).
\end{itemize}

\section*{Acknowledgements}
\rs{Chiara Lelli was partially supported by
Grant 5 per Mille Junior 2009
"Computational Models for Heterogeneous Media. Application
to Micro Scale Analysis of Tissue-Engineered Constructs"
CUPD41J10000490001 Politecnico di Milano.
Manuela T. Raimondi has received
funding from the European Research Council (ERC) under the European
Union's Horizon 2020 research and innovation program (Grant
Agreement No. 646990-NICHOID). These results reflect only the
author's view and the agency is not responsible for any use that may be
made of the information contained.}

\bibliographystyle{plain}      
\bibliography{PAPERMB}   

\begin{thebibliography}{10}

\bibitem{AmbrosiPreziosi}
D.~Ambrosi and L.~Preziosi.
\newblock On the closure of the mass balance models for tumor growth.
\newblock {\em Math. Models Methods Appl. Sci.}, 12:737--754, 2002.

\bibitem{Araujo}
R.~P. Araujo and D.~L.~Sean McElwain.
\newblock A mixture theory for the genesis of residual stresses in growing
  tissues {I}: a general formulation.
\newblock {\em SIAM Journal of Applied Mathematics}, 65(4):1261--1284, 2005.

\bibitem{Bader2011}
D.~L. Bader, D.~M. Salter, and T.~T. Chowdhury.
\newblock Biomechanical influence of cartilage homeostasis in health and
  disease.
\newblock {\em Arthritis}, page Article ID 979032, 2011.

\bibitem{Barry}
S.~I. Barry and G.~N. Mercer.
\newblock Flow and deformation in poroelasticity {- I} unusual exact solutions.
\newblock {\em Mathematical and Computer Modelling}, 30:23--29, 1999.

\bibitem{biot1}
M.~A. Biot.
\newblock General {T}heory of {T}hree-{D}imensional {C}onsolidation.
\newblock {\em J. Appl. Phys.}, 12(2):155--164, 1941.

\bibitem{Moreo3D}
C.~Borau, R.~D. Kamm, and J.~M. Garcia-Aznar.
\newblock Mechano-sensing and cell migration: a {3D} model approach.
\newblock {\em Phys. Biol.}, 8:1--13, 2011.

\bibitem{ByrnePreziosi2003}
H.~Byrne and L.~Preziosi.
\newblock {Modelling solid tumour growth using the theory of mixtures}.
\newblock {\em Math Med Biol}, 20(4):341--366, December 2003.

\bibitem{Causin}
P.~Causin, R.~Sacco, and M.~Verri.
\newblock A multiscale approach in the computational modeling of the
  biophysical environment in artificial cartilage tissue regeneration.
\newblock {\em Biomechanics and Modeling in Mechanobiology}, 12(4):763--780,
  2013.

\bibitem{Cheng}
G.~Cheng, B.~B. Youssef, P.~Markenscoff, and K.~Zygourakis.
\newblock Cell population dynamics modulate the rates of tissue growth
  processes.
\newblock {\em Biophys J.}, 90(3):713--724, 2006.

\bibitem{Chung}
C.~A. Chung, Tze-Hung Lin, Shih-Di Chen, and Hsing-I Huang.
\newblock Hybrid cellular automaton modeling of nutrient modulated cell growth
  in tissue engineering constructs.
\newblock {\em J. Theor. Biol.}, 262:267--278, 2010.

\bibitem{Coller}
H.~A. Coller, L.~Sang, and J.~M. Roberts.
\newblock A new description of cellular quiescence.
\newblock {\em PlOs Biol.}, 4(3):e83, 2006.

\bibitem{Contois}
D.~E. Contois.
\newblock Kinetics of bacterial growth: relationship between population density
  and specific growth rate of continuous cultures.
\newblock {\em J. Gen. Microbiol.}, 21:40--50, 1959.

\bibitem{Coussy}
O.~Coussy.
\newblock {\em Poromechanics}.
\newblock John Wiley {\&} Sons, 2004.

\bibitem{DiMilla1991}
P.A. DiMilla, K.~Barbee, and D.A. Lauffenburger.
\newblock Mathematical model for the effects of adhesion and mechanics on cell
  migration speed.
\newblock {\em Biophysical Journal}, 60(1):15--37, 1991.

\bibitem{Ducrot}
A.~Ducrot, F.~Le Foll, P.~Magal, H.~Murakawa, J.~Pasquier, and G.~F. Webb.
\newblock An in vitro cell population dynamics model incorporating cell size,
  quiescence, and contact inhibition.
\newblock {\em Math. Mod. Meth. Appl. S.}, 21:871--892, 2011.

\bibitem{Freed}
L.~E. Freed, I.~Martin, and G.~Vunjak-Novakovic.
\newblock Frontiers in tissue engineering: In vitro modulation of
  chondrogenesis.
\newblock {\em Clin. Orthop.}, 367S:S46--S58, 1999.

\bibitem{Freed2000}
L.E. Freed and G.~Vunjak-Novakovic.
\newblock Tissue engineering bioreactors.
\newblock In R.~P. Lanza, R.~Langer, and J.~Vacanti, editors, {\em Principles
  of tissue engineering}. Academic Press, San Diego, 2000.

\bibitem{frijns}
A.~J.~H. Frijns.
\newblock {\em A four-component mixture theory applied to cartilaginous
  tissues: {N}umerical modelling and experiments}.
\newblock ProQuest LLC, Ann Arbor, MI, 2000.
\newblock Thesis (Dr.ir.)--Technische Universiteit Eindhoven (The Netherlands).

\bibitem{Galbusera2008}
F.~Galbusera, M.~Cioffi, and M.~T. Raimondi.
\newblock An in silico bioreactor for simulating laboratory experiments in
  tissue engineering.
\newblock {\em Biomedical Microdevices}, 10(4):547--554, 2008.

\bibitem{Galbusera2007}
F.~Galbusera, M.~Cioffi, M.~T. Raimondi, and R.~Pietrabissa.
\newblock Computational modelling of combined cell population dynamics and
  oxygen transport in engineered tissue subject to interstitial perfusion.
\newblock {\em Computer Methods in Biomechanics and Biomedical Engineering},
  10(4):279--287, 2007.

\bibitem{Klisch_intrinsic_incompressibility}
S.~M. Klisch.
\newblock Internally constrained mixtures of elastic continua.
\newblock {\em Mathematics and Mechanics of Solids}, 4:481--498, 1999.

\bibitem{Klisch1}
S.~M. Klisch, S.~S. Chen, R.~L. Sah, and A.~Hoger.
\newblock A growth mixture theory for cartilage with application to
  growth-related experiments on cartilage explants.
\newblock {\em J. Biomech. Eng.}, 125:169--179, 2003.

\bibitem{Klisch0}
S.~M. Klisch, T.~J.~Van Dyke, and A.~Hoger.
\newblock A theory of volumetric growth for compressible elastic biological
  materials.
\newblock {\em Math. Mech. Solids}, 6:551--575, 2001.

\bibitem{Klisch}
S.~M. Klisch, R.~L. Sah, and A.~Hoger.
\newblock A cartilage growth mixture model for infinitesimal strains: solutions
  of boundary-value problems related to in vitro growth experiments.
\newblock {\em Biomech. Model. Mechanobiol.}, 3:209--223, 2005.

\bibitem{Langer1993}
R.S. Langer and J.P. Vacanti.
\newblock Tissue engineering.
\newblock {\em Science}, 260(5110):920--926, 2004.

\bibitem{Lemon2007}
G.~Lemon and J.~R. King.
\newblock Multiphase modelling of cell behaviour on artficial scaffolds:
  effects of nutrient depletion and spatially nonuniform porosity.
\newblock {\em Mathematical Medicine and Biology}, 24:57--83, 2007.

\bibitem{Lemon2006}
G.~Lemon, J.~R. King, H.~M. Byrne, O.~E. Jensen, and K.~M. Shakesheff.
\newblock Mathematical modelling of engineered tissue growth using a multiphase
  porous flow mixture theory.
\newblock {\em Journal of Mathematical Biology}, 52:571--594, 2006.

\bibitem{Maini2002}
P.K. Maini, J.~A. Sherratt, and L.~Olsen.
\newblock Mathematical models for cell-matrix interactions during dermal wound
  healing.
\newblock {\em Int. J. Bif. Chaos}, 12(9):2021--2029, 2002.

\bibitem{Martin2004}
I.~Martin, D.~Wendt, and M.~Heberer.
\newblock The role of bioreactors in tissue engineering.
\newblock {\em Trends Biotechnol.}, 22(2):80--86, 2004.

\bibitem{Moreo2}
P.~Moreo, E.~A. Gaffney, J.~M. Garcia-Aznar, and M.~Doblar\'{e}.
\newblock On the modelling of biological patterns with mechanochemical models:
  insights from analysis and computation.
\newblock {\em Bull. Math. Biol.}, 72:400--431, 2010.

\bibitem{Moreo1}
P.~Moreo, J.~M. Garcia-Aznar, and M.~Doblar\'{e}.
\newblock Modeling mechanosensing and its effect on the migration and
  proliferation of adherent cells.
\newblock {\em Acta Biomater.}, 4:613--621, 2007.

\bibitem{Nava2012}
M.M. Nava, M.T. Raimondi, and R.~Pietrabissa.
\newblock Controlling self-renewal and differentiation of stem cells via
  mechanical cues.
\newblock {\em Journal of Biomedicine and Biotechnology}, 2012:12, 2012.

\bibitem{Nava2014}
M.M. Nava, M.T. Raimondi, and R.~Pietrabissa.
\newblock Bio-chemo-mechanical models for nuclear deformation in adherent
  eukaryotic cells.
\newblock {\em Biomechanics and Modeling in Mechanobiology}, 13(5):929--943,
  2014.

\bibitem{Nikolaev}
N.I. Nikolaev, B.~Obradovic, H.K. Versteeg, G.~Lemon, and D.J. Williams.
\newblock A validated model of gag deposition, cell distribution, and growth of
  tissue engineered cartilage cultured in a rotating bioreactor.
\newblock {\em Biotechnol Bioeng.}, 105(4):842--853, 2010.

\bibitem{Osborne}
J.~M. Osborne, R.~D. O'Dea, J.~P. Whiteley, H.~M. Byrneand, and S.~L. Waters.
\newblock The influence of bioreactor geometry and the mechanical environment
  on engineered tissues.
\newblock {\em J. Biomech. Eng.}, 132 (5):051006, 2010.

\bibitem{Murray}
G.~F. Oster, J.~D. Murray, and A.~K. Harris.
\newblock Mechanical aspects of mesenchymal morphogenesis.
\newblock {\em J. Embryol. exp. Morph.}, 78:83--125, 1983.

\bibitem{Tosin}
L.~Preziosi and A.~Tosin.
\newblock Multiphase modelling of tumour growth and extracellular matrix
  interaction: mathematical tools and applications.
\newblock {\em Journal of Mathematical Biology}, 58:625--656, 2009.

\bibitem{Raimondi2005}
M.~T. Raimondi, F.~Boschetti, F.~Migliavacca, M.~Cioffi, and G.~Dubini.
\newblock Micro fluid dynamics in three-dimensional engineered cell systems in
  bioreactors.
\newblock In N.~Ashammakhi and R.~L. Reis, editors, {\em Topics in Tissue
  Engineering}, volume~2, chapter~9. 2005.

\bibitem{Raimondi2008}
M.~T. Raimondi, G.~Candiani, M.~Cabras, M.~Cioffi, M.~Lagan\`{a}, M.~Moretti,
  and R.~Pietrabissa.
\newblock Engineered cartilage constructs subject to very low regimens of
  interstitial perfusion.
\newblock {\em Biorheology}, 45:471--8, 2008.

\bibitem{raimondipellets2011}
M.T. Raimondi, E.~Bonacina, G.~Candiani, M.~Lagan\`a, E.~Rolando, G.~Tal\`a,
  D.~Pezzoli, R.~D'Anchise, R.~Pietrabissa, and M.~Moretti.
\newblock Comparative chondrogenesis of human cells in a {3D} integrated
  experimental–computational mechanobiology model.
\newblock {\em Biomechanics and Modeling in Mechanobiology}, 10(2):259--268,
  2011.

\bibitem{Raimondi2006a}
M.T. Raimondi, M.~Moretti, M.~Cioffi, C.~Giordano, F.~Boschetti, K.~Lagan\`a,
  and R.~Pietrabissa.
\newblock The effect of hydrodynamic shear on {3D} engineered chondrocyte
  systems subject to direct perfusion.
\newblock {\em Biorheology}, 43(3-4):215--222, 2006.

\bibitem{Tew2001}
Tew S., Redman S., Kwan A., Walker E., Khan I., Dowthwaite G., Thomson B., and
  Archer C.W.
\newblock Differences in repair responses between immature and mature
  cartilage.
\newblock {\em Clin Orthop Relat Res.}, 391 Suppl:S142--52, 2001.

\bibitem{Sacco}
R.~Sacco, P.~Causin, P.~Zunino, and M.~T. Raimondi.
\newblock A multiphysics/multiscale {2D} numerical simulation of scaffold-based
  cartilage regeneration under interstitial perfusion in a bioreactor.
\newblock {\em Biomech Model Mechanobiol}, 10(4):577--589, 2011.

\bibitem{Sengers}
B.~G. Sengers, C.~W.~J. Oomens, and F.~P.~T Baaijens.
\newblock An integrated finite-element approach to mechanics, transport and
  biosyntheis in tissue engineering.
\newblock {\em Journal of Biomechanical Engineering}, 126(1):82--91, 2004.

\bibitem{Sengers2}
B.~G. Sengers, M.~Taylor, C.~P. Please, and R.~O.C. Oreffo.
\newblock Computational modelling of cell spreading and tissue regeneration in
  porous scaffolds.
\newblock {\em Biomaterials}, 28:1926--1940, 2007.

\bibitem{SengersPhD}
BG~Sengers.
\newblock {\em Modeling the development of tissue engineered cartilage}.
\newblock PhD thesis, Department of Biomedical Engineering, Technische
  Universiteit Eindhoven, 2005.

\bibitem{sokolnikoff1956}
I.~S. Sokolnikoff.
\newblock {\em Mathematical theory of elasticity}.
\newblock Mc Graw-Hill, New York, 1956.

\bibitem{Tosin2}
A.~Tosin.
\newblock Multiphase modeling and qualitative analysis of the growth of tumor
  cords.
\newblock {\em Netw. Heterog. Media}, 3:43--83, 2008.

\bibitem{Please}
A.~J. Trewenack, C.~P. Please, and K.~A. Landman.
\newblock A continuum model for the development of tissue-engineered cartilage
  around a chondrocyte.
\newblock {\em Mathematical Medicine and Biology}, 26:241--262, 2009.

\bibitem{Novakovic}
G.~Vunjak-Novakovic, B.~Obradovic, I.~Martin, P.~M. Bursac, R.~Langer, and
  L.~E. Freed.
\newblock Dynamic cell seeding of polymer scaffolds for cartilage tissue
  engineering.
\newblock {\em Biotechnology Progress}, 14(2):193--202, 1998.

\bibitem{whitaker1999}
S.~Whitaker.
\newblock {\em The method of volume averaging. Theory and application of
  transport in porous media}.
\newblock Kluwer Academic Publishers, 1999.

\bibitem{Wood2002}
B.~D. Wood, M.~Quintard, and S.~Whitaker.
\newblock Calculation of effective diffusivities for biofilms and tissues.
\newblock {\em Biotech. and Bioeng.}, 77(5):495--514, 2002.

\end{thebibliography}

\end{document}